\documentclass[a4paper,12pt]{article}
\usepackage{epsfig}
\usepackage{amssymb}
\begin{document}
\thispagestyle{empty}

%____________________________________________
%\newcommand{\JP}    {jet probability}
\newcommand{\etal}  {{\it{et al.}}}  % must have all the {}
\def\Journal#1#2#3#4{{#1} {\bf #2}, #3 (#4)}
\def\PRD{Phys.\ Rev.\ D}
\def\NIMA{Nucl.\ Instrum.\ Methods A}
\def\PRL{Phys.\ Rev.\ Lett.\ }
\def\PLB{Phys.\ Lett.\ B}
\def\EPJ{Eur.\ Phys.\ J}
\def\IEEETNS{IEEE Trans.\ Nucl.\ Sci.\ }
\def\CPCD{Comput.\ Phys.\ Commun.\ }
%____________________________________________

%\hfill {\LARGE\bf DRAFT}
%\smallskip

%\hfill {\large\bf \today}
\bigskip
%\bigskip

%{\LARGE\bf
{\Large\bf
\begin{center}
Random fluctuation walk in the boson star formation process
\end{center}
}
\vspace{0.1 cm}
%\vspace*{\fill}

\begin{center}
{ G.A. Kozlov  }
\end{center}
%\vspace{1 cm}
%\vspace*{\fill}
%\date
\begin{center}
\noindent
 { Bogolyubov Laboratory of Theoretical Physics\\
 Joint Institute for Nuclear Research,\\
 Joliot Curie st., 6, Dubna, Moscow region, 141980 Russia  }
\end{center}
%\vspace*{\fill}
\vspace{0.1 cm}

 %\section*{Abstract}
 \begin{abstract}
 \noindent
 { We construct the mechanism of formation of the scalar boson star (BS) having a hierarchy of self-similar "lumps" of dark matter (DM) built into the probability distribution function of allowed steps.
It can ensure to study the dynamics of the BS formation through the "cross-over" between the free scalar DM and the compact condensate (CC) phase in one-dimensional model of the random fluctuation walk. The main inputs are the random fluctuating weight $\lambda < 1$, the fundamental fluctuating length $\xi (\lambda)$ and the parameter $a >1$ of the spatial separation between the DM "lumps" inside the BS. The solution to the functional equation for the characteristic probability to find the free scalar DM and the CC phase is obtained.
The time evolution of the probability to form the DM "lump" (growth of the BS) is presented with the power-law cascades expression depending on $\lambda$ and $a$.
The special cases for $\lambda$ in terms of the energy densities of the BS ($\rho_\star$)  and the DM ($\rho_\odot$), as well as the Bose-Einstein correlation function for two scalar DM particles, are considered. The lower bounds on the critical temperature at the "cross-over" depending on $\rho_\odot$,  $\rho_\star$, the average multiplicities of the scalar DM particles and the parameters related to the chaotically behaviour of the particles are obtained.}

%\vspace*{\fill}

\end {abstract}

%\vspace*{\fill}

%\newpage
%\tableofcontents

%\newpage
%\section{Introduction}

\bigskip

\section{Introduction}
%\label{sec:level1}First-level heading:\protect\\ The line
%break was forced \lowercase{via} \textbackslash\textbackslash}
The modern scenarios to form the macroscopic Bose-like objects in the Universe broach the question concerning the role of the scalar (dilaton) field and its influence on development of cosmological  inhomogeneities. 
The latter are related to the scalar boson star (BS) as the gravitationally bound field state [1-3] composed dominantly of the scalar dark matter (DM) fields $\phi$ that may be lighter than the spin-1/2 DM particle $\chi$ treated as a Dirac particle. 
%The $\phi$ may be the dominant product of the $\bar\chi\chi$ annihilation. 
The $\mathbb{Z}_2$ symmetry protects $\chi$ as a stable particle. One of the  physically promising $\bar\chi\chi$ annihilation channels is into two $\phi$ particles via exchanging $\chi$ with the interaction of the form $\sim (1 + \phi/\phi_0)m_\chi \bar\chi\chi$, where $\phi_0\sim O(1\,TeV)$ [4] is the scale invariance breaking mass, $m_\chi$ is the thermally produced  $\chi$-DM mass of the order $\leq O(100\, TeV) $[5-7]. 
The evolution of Dirac and Bose particles has been studied in the framework of multi-particle correlation and distribution functions to predict the size of the finite-temperature phase transition domain [8].
We admit the thermal bath, where the scalar DM are bounded dominantly  by the gravitational  forces and undergo the interactions with the gauge fields and the Higgs bosons [9].

The behaviour of the free scalar DM particles in the approximate scale-invariant symmetry world and the transition to the state being gravitationally bound in the BS at some "critical point", has an interest for many decades. The $\phi$ particles correlate to each other in the space-time upon produced due to $\bar\chi\chi$ annihilation. The study of the scalar DM condensate can be done through the Bose-Einstein correlations (CBE) of identical particles. The $\phi$ particles can undergo the correlations in some finite space and then gravitationally clustered into macroscopic objects, e.g., the "lumps" [10] distributed inside the BS with respect to CBE principles. The CBE can allow to define the spatial domains of the sources (the "lumps") where the $\phi$ particles are born of and then be captured by the gravitational forces. For our aim, the usage of two-particle CBE function $C_2(q,\lambda_{ch})$ with $q$ and $\lambda_{ch}$ being the relative momentum between two particles and the coherence function, respectively, can be an effective tool to an understanding of the dynamics in the BS formation. 
%through the spatial correlations between the particles. 
The BS linear size, $\sim R_\star$, is characterised by many stochastic scales $L_{st}$, $R_\star \sim \sum_i L_{st,i}$, where each $L_{st,i}$ defines an effective size of the DM "lump". The $L_{st}$ depends on the temperature $T$ to the bath, the $q$ and can undergo the influence of the random fields parametrised by the so-called function as  the chaoticity strength $\nu$ which goes to zero when the "cross-over" is approached.  
The latter means the transition between the free DM fields to the Bose-Einstein condensation (BEC) phase under the gravitational interactions.
The strength $\nu$ is the effective number of $\phi$ particles with the mass $m_\phi$ in the plane phase space with the size $\sim L_{st}$,
$$ \nu(m_\phi, T) \sim \frac{\hat n(\omega, T)}{\hat m^2_\phi (\omega, T) L_{st}^2}, $$
where $\hat m_\phi (\omega, T) = m_\phi\, \hat n(\omega, T)$, $\hat n(\omega, T) = {\left [ e^{(\omega - \mu)/T} -1 \right ]}^{-1}$;
$\omega$  is the energy of $\phi$-particle with momentum $p = (\omega, \vec p)$ in the bath with statistical equilibrium; $\mu$ is the chemical potential. The consideration is correct for large occupation numbers  $\hat n(\omega, T)$ where the DM particles are light. Actually, $\nu\rightarrow 0$ as $\hat n(\omega, T)\rightarrow \infty$ at the critical temperature $T_c$. %Here, we have an ensemble of light SDM particles (small $m_\phi$) with large $\hat n(\omega, T)$ which thermalises into a BEC with the peak of the BS formation when the "chaoticity" strength $\nu$ goes to zero. 
The function $\lambda_{ch} (\nu)$ represents the degrees of the coherence and the chaoticity to an emission effect in the particle source: $\lambda_{ch} (\nu)\sim \gamma (\omega, T)/ (1 + \nu)^2$, where $\gamma (\omega, T) = \hat n^2 (\bar\omega)/\left [ \hat n(\omega) \hat n(\omega^\prime)\right ]$  defines the quantum thermal properties of the particles source, $\bar\omega = (\omega +\omega^\prime)/2$.
The  $\lambda_{ch}$ at given $T$ runs from $\lambda_{ch} = 0$ (a fully coherent phase) to $\lambda_{ch} = 1$ (a fully chaotic phase at the"cross-over"). The details related to $C_2(q,\lambda_{ch})$ function can be found in [11]. 
Here, we have an ensemble of light scalar DM particles (small $m_\phi$ with large $\hat n(\omega, T))$ which thermalises into a BEC with the peak of the BS formation when the chaoticity strength $\nu$ goes to zero. 
To search the early stage of the BS formation, we use the approach to "random fluctuation walks" (RFW) with respect to the chaoticity in correlations of identical $\phi$ particles, as well as the energy density of the DM and the characteristic effective scales of the BS. 
%The PT means the transition between the free SDM fields to the Bose-Einstein condensation (BEC) phase under the gravitational interactions.
The inputs of the BS formation are the initial conditions $R_\star >> {\left (m_\phi\cdot v\right)}^{-1}$, $\tau_{cond} >> {\left (m_\phi\cdot v^2 \right)}^{-1}$ with $\tau_{cond}$ being the time of the condensation and kinetic variable $v$ is the velocity. 
%The large enough DM density may be a source of a (local) spontaneous breaking space parity of $\bar\chi\chi$ annihilation into the SDM fields through the BEC. 
%The breaking of space parity have already been investigated in [11] in the form of chiral magnetic effect at  non-central heavy-ion collisions, and in [12] through the anomaly %yield of leptonic pairs. 
The mechanism of an appearance of the scalar DM condensate inside the hot macroscopic object at hight DM energy density  has not been understand yet. It might be one of the goals to search program to the BS formation. 
The "adiabatic" approach to the self-similar solution of the kinetics problems of the BS growth in the bath of the gravitationally interacting particles has been studied in [12].
The DM energy density $\rho_\odot = 0.43\,GeV\, cm^{-3}$ in the vicinity of the Solar system [13] is very large and the BEC of the scalar DM with the energy density  $\rho < <\rho_\odot$ can appear in the cosmological structures with the gravitational interaction of the scalar DM in the so-called "dynamical approach" to the cosmological scale, the Planck mass $M_{Pl}$ [14]. This means the BS may occur in the homogeneous cosmological structure with some probabilities that may be realised  through the RFW formalism with the self-similar solution of the dynamical equation.  

\section{RFW}
We consider an effective theory in what follows that the Universe at its early stage was driven by the scalar DM fields that were minimally coupled to gravity (GR) in the sense of dynamical fields. The main contribution to the BS formation comes from the strong DM scalar sector where $M_{Pl}^2 << {\vert \phi\vert}^2$.  The Lagrangian density (LD) is
%has the form 
\begin{equation}
\label{e1}
\frac{L}{\sqrt {-g}} = \frac{1}{2}  R \,\zeta_\phi {\vert \phi\vert}^2 - \frac{1}{2} g^{\mu\nu}\partial_\mu\phi\,\partial_\nu\phi^\star + L_D,
\end{equation}
where the first term is the Einstein-Hilbert action for GR with $R$ being the Ricci scalar for background metric $g_{\mu\nu}$; the parameter $\zeta_\phi$ defines the finite regime of validity, i.e., the cutoff energy scale at which the effective theory breaks down,
%the cutoff of the theory, 
$\Lambda_{cut}\sim M_{Pl}/\sqrt{\zeta_\phi}$; $L_D$ is the LD containing the interactions of the complex fields $\phi$ with the vector (gauge) fields against the scale and gauge invariances [9].  The (\ref{e1}) obeys the scale invariance with $M_{Pl}^2 \sim\zeta_\phi {\vert \phi\vert}^2$. 
For LD with the potential $\sim (1/2)m_\phi^2 {\vert\phi\vert}^2 + (1/4)\lambda_4  {\vert\phi\vert}^4 $ the astrophysical stage with the Hubble constant $ H << m_\phi$ ($\hbar = c =1$) corresponds to the dust-like expansion stage at which the formation of the inhomogeneities are possible (see, e.g., [15] and the refs. therein).
The "cross-over" between the free scalar DM particles and the ones being in the condensate (inside the star), is related with the fluctuations of $\phi$ excitations  with $T$, $\rho$ and the scalar DM density number $n$.
The BS itself is considered as the scalar bound state occupied by a lot number of the local DM "lumps" (composed of the scalar fields) separated by the finite spacing $\sim a/\Lambda_{cut}$.
For simplicity, we consider the $D=1$ spatial dimension $x$-oriented axis model to the scalar DM fluctuations where the probability of occupancy of the $x$th "lump" after $n$ steps inside the BS is 
\begin{equation}
\label {e2}
\Pi_{n^\prime} (x;\lambda) = \sum_{x^\prime} P(x-x^\prime; \lambda) \Pi_n (x^\prime;\lambda), \,\, -\infty < x^\prime < +\infty,
\end{equation}
where $n^\prime > n$, $0 <\lambda <1$ is the random fluctuation weight that can define the number of $\phi$-particles in a "lump". The probability density in (\ref{e2}) for a step $x$ is governed by distribution function $P(x;\lambda)$ in a symmetric random walk, $P(x;\lambda) = P(-x;\lambda)$, with a step of length $\sim a^j/\Lambda_{cut}$, 
\begin{equation}
\label {e03}
%P (x;\lambda) = p(\lambda)\sum_{j=0}^\infty \lambda^j \frac{1}{2\sqrt{\pi\theta}}
%\left [e^{{-\left (y_{+}^j\right )}^2/(4\theta)} + e^{{-\left (y_{-}^j\right )}^2/(4\theta)}\right], 
P (x;\lambda) = \frac {p(\lambda)}{2\sqrt{\pi\theta}} \sum_{j=0}^\infty \lambda^j \left [e^{{-\left (y_{+}^j\right )}^2/(4\theta)} + e^{{-\left (y_{-}^j\right )}^2/(4\theta)}\right], 
\end{equation}
where $\lambda$ is related either to $\rho$, or to $C_2 (q, \lambda_{ch})$; $y_{\pm}^j = x\Lambda_{cut} \pm \Delta a^j$ with $a >1$ and $\Delta > 0$; $\theta = l\Lambda_{cut}$, $l$ is the fundamental minimal scale, the "lump" spacing. The  Gaussian form (\ref{e03}) corresponds to a spreading cluster in the "lump" and ensures that as $j\rightarrow\infty$, the random walk spreads out to occupy all sites in the "lumps" with slowly varying homogeneous distribution. The factor $p(\lambda)$ will be defined later. The (\ref{e03}) allows to form a DM 'lump" before moving an order of magnitude further away in beginning to form a new DM "lump". The mean-square displacement $\langle x^2\rangle \sim \Delta^2 \sum_{m=0}^\infty {\left (a^2 \lambda\right)}^m$ is infinite if $\lambda > a^{-2}$. In field theory, there is a continuum limit in the space-time $\langle x^2\rangle\rightarrow\infty$ per step, that can lead to $\Pi (x,t;\lambda)$ distribution a time $t$ after the DM "lump" walk formation begins and the RFW be transient. The stars of the size $\sim R_\star$ are wide compared to the inverse mass of the scalar DM, $r_\star \sim m_\phi^{-1}$.
We define the variable $a$ in terms of the spatial separation of the DM "lumps", $a =\Delta_\star r_\star \phi_0$ where $\Delta_\star r_\star$ is controlled by the dimensionless parameter [3] 
$g = \lambda_4 M_{Pl}^2/(4\pi m_\phi^2)$. Here, $\Delta_\star = 1$ for the DM "lump", while $\Delta_\star\sim \sqrt g$ for the star; $r_\star << R_\star \sim \sqrt g/m_\phi$ and $\lambda_4$ is the repulsive self-interaction coupling constant in potential with the term $\sim \lambda_4 {\vert\phi\vert}^4$ (see also [16]).  
%$\Delta_\star r_\star $
%We define $a$ in terms of the spatial separation of the "lumps", $r_\star \sim m_\phi^{-1} << R_\star \sim \sqrt g_4/m_\phi$, where $g_4$ is the dimensionless parameter of the order $\sim O\left %%( \lambda_4\,M_{Pl}^2/m_\phi^2\right )$ with $\lambda_4$ being the repulsive self-interacting coupling constant in $\sim \lambda_4 \phi^4$ term of the LD [15]. 
The parameter $a$ increases with $\Delta_\star r_\star$ and has the  form $a = A\, \phi_0\, T_c \,n^{-2/3}$ when the star formation is completed with
% Here, $\Delta_\star$ is the positive number ($\Delta_\star = 1$ for the DM "lump", while $\Delta_\star \sim\sqrt {g_4}$ for the whole star);  
$A = \zeta ^{2/3} (3/2)/(2\pi)$, where  $\zeta(x)$ is the Riemann's  Zeta-function. The density number $n$ is estimated from the relation $\rho_\odot = M_\star n_\star + m_\phi n (1 + \delta_h)$, where  $M_{\star}$ and $n_\star$ are the mass and the density number of the BS. Both, $n_\star$ and the Higgs contribution parameter $\delta_h$ are estimated in [9]. The critical temperature $T_c$, related to the "cross-over", 
%from the free SDM (FM) to the compact condensate (CC) bounded inside the BS, 
may be defined from the condition that the series (\ref{e2})  converges. In fact, the (\ref{e2}) is the probability density for a step $x\cdot \Lambda_{cut}$ which exhibits the DM clustering in the DM "lump" with the step length $\sim a^j/\Lambda_{cut}$ and the probabilities $\lambda$ corresponding to $a^j/\Lambda_{cut}$.  In the limit $l\rightarrow 0$, the distribution (\ref{e2})  becomes a delta-shaped sequence, where $p(\lambda) = (1-\lambda)/2$. The values $\lambda >1$ are excluded because of the normalisation condition $2p (1 + \lambda + ... + \lambda^j + ...) =1$. If $ \lambda\rightarrow 1$, the $P(x;\lambda\rightarrow 1)$ is rather broad and slowly retarding that may indicate the vicinity of the "cross-over". Contrary to that,  $P(x;\lambda\rightarrow 0)\rightarrow 1/2$ is trivial, so one has the range where the "cross-over" can be found at appropriate $\rho$ and/or the $C_2(q,\lambda_{ch})$ function. In order to smooth the particularities in (\ref{e2}), we use the characteristic function $G(k;\lambda)$, the Fourier transformation of  $P(x;\lambda)$
\begin{equation}
\label{e3}
G(k;\lambda) = p(\lambda) \sum_{j = 0}^\infty \lambda^j\,\cos \left (\frac{k}{\Lambda_{cut}}\,a^j\right).
\end{equation}
The $k\rightarrow 0$ behaviour of (\ref{e3}) will determine the differential equation for $\Pi(x,t;\lambda)$ in the continuum limit and the transience of the walk from the free scalar DM to BEC.
The non-analytic behaviour of  $G(k;\lambda)$  in (\ref{e3}) at $k = 0$ can be exhibited if we use the inverse Mellin transformation with respect to $k$,
%$$ G(k;\lambda) =  \frac{p(\lambda)}{2\pi i} \int_{\kappa - i\infty}^{\kappa + i\infty} \Gamma (y)\,\cos \left (\frac{1}{2} \pi y\right ) {\left (\frac{\Lambda_{cut}}{k}\right )}^y
%\sum_{j=0}^\infty \lambda^j\frac{1}{(a^j)^y} dy, $$
$$ G(k;\lambda) =  \frac{p(\lambda)}{2\pi i} \int_{\kappa - i\infty}^{\kappa + i\infty} \Gamma (y)\,\cos \left (\frac{\pi y}{2}\right ) {\left (\frac{\Lambda_{cut}}{k}\right )}^y
\sum_{j=0}^\infty \frac{\lambda^j}{(a^j)^y} dy, $$
%[1 + \lambda {\left (\frac{\Lambda_{cut}}{a}\right )}^y\right ] dy,$$
where $ 0 < \kappa = Re \,y < 1$.
%for all the values $ \lambda < A\,\phi_0\, T_c /\left (n^{2/3}\,\Lambda_{cut}\right ).$  
For every value of dimensionless $ z= k/(\pi\,\Lambda_{cut})$  with $ z\rightarrow z + \Delta z$ one has [17] $\tilde G (z + \Delta z; \lambda) = \tilde G (z; \lambda) + O\left ( {\vert \Delta z\vert}^\beta\right )$, where $\tilde G(z;\lambda) =  G(z;\lambda)/p(\lambda)$ is the Weierstrass's function and $\beta = \left [ \ln \left (1/\lambda \right )/\ln a\right ] < 1$. 

Any observable associated with (\ref{e3}) undergoes the RFWs towards the "cross-over". The random walk is characterised by the fundamental length $\xi$,
 \begin{equation}
\label{e4}
\vert \xi^{2S}(\lambda)\vert  = p(\lambda) \sum_{j = 0}^\infty \lambda^j\,\left (\frac{a^j}{\Lambda_{cut}}\right)^{2S},
\end{equation}
which increases when $n\rightarrow 0$ at $T_c\rightarrow\infty$. The result  (\ref{e4}) is the derivative of (\ref{e3}) on $k$ to the order $2S$ ($S =1,2,...$) at $k = 0$. The fundamental length is sensitive to the values $\lambda \cdot a^{2S}$. To clarify this sensitivity, one can move to Taylor's series (instead of (\ref{e3})),
 \begin{equation}
\label{e5}
G(k;\lambda) = 1 +   \sum_{S = 1}^\infty   \frac{1}{\left ( 2S \right )!}\,i^{2S}\, \xi^{2S}( \lambda) k^{2S},
\end{equation}
which has the convergence when the critical temperature is restricted by $T_c < \lambda^{1/(2S)} n^{2/3}/(A\,\phi_0)$ for any random weight $\lambda < 1$ when 
$\vert \xi^{2S}(\lambda)\vert \cdot \Lambda_{cut}^{2S} $ will be finite
 for all the orders $2S$ $(S = 1,2,...)$ with $\Lambda_{cut}\sim O(M_{Pl})$, if $\zeta_\phi\sim O(1)$. The  $\xi(\lambda)$ increases (as $\lambda\rightarrow 1$) up to divergence that defines the "cross-over". One can find an infinite number of divergent terms starting from (\ref{e3}), hence neither (\ref{e3}), nor the (\ref{e5}) do not suitable to describe the arbitrary phases of the DM in the range $0 < \lambda < 1$. For an arbitrary set of the variable $\lambda$ and the parameter $a$ acceptable both for free DM and the compact condensate (CC) phases, we use the functional linear non-homogeneous equation for the probability to find DM in the free state and in the CC phase, 
 \begin{equation}
\label{e6}
G(k;\lambda) =  G_{DM}(k;\lambda) + G_{CC}(a\cdot k;\lambda),
\end{equation}
where the non-analytical part at $k\rightarrow 0$ may be clarified. In (\ref{e6}),
\begin{equation}
\label{e7}
G_{DM}(k;\lambda)  = p(\lambda)\,\cos \left ( k/\Lambda_{cut}\right),
\end{equation}
while 
\begin{equation}
\label{e8}
G_{CC}(a\cdot k;\lambda)  = \lambda\,  G(a\cdot k;\lambda).
\end{equation} 
The  (\ref{e7}) is the only term that gives the contribution to the DM, while the CC phase is provided by (\ref{e8}).
%for $ 1 > \lambda \geq a^{-2}$.
Note, that the probability function (\ref{e7}) is the non-homogenious $k$-dependent function which is regular in the vicinity of $k = 0$ for all the values in the range $0 < \lambda < 1$. The (\ref{e8}) disappears as $\lambda\rightarrow 0$ ($n\rightarrow 0$ and $T_c\rightarrow\infty$) and the DM in the free states can exist only. Actually, in the interval $0 < \lambda < 1$ the "cross-over" may happen many times as the cascade with definite $a$ against the boundary condition $\lambda a^{2S} < 1. $ Once the BS appears, the probability $G(k;\lambda)$ divides between the free DM, $G_{DM}$, and the star object as excited bound state with BEC, $G_{CC}$. The condition for condensation is $\lambda > a^{-2}$.  

The solution for the regular part of (\ref{e5}) for DM is 
  \begin{equation}
\label{e9}
G_{DM}\left [k;\lambda (S)\right ] = 1 +   \sum_{s = 1}^{S-1} (-1)^s  \frac{1}{\left ( 2s \right )!}\, \xi^{2s}\left [ \lambda (s)\right ] k^{2s},
\end{equation}
which is regular at $0 < \lambda < 1$ with finite $S$. Note, that $S\rightarrow\infty$ as $\lambda\rightarrow 0$. The rest term is singular
 \begin{equation}
\label{e10}
G_{CC}\left [k;\lambda (S)\right ] = 1 +   \sum_{s = S }^\infty  (i)^{2s} \frac{1}{\left ( 2s \right )!}\, \xi^{2s}\left [ \lambda (s)\right ] k^{2s}
\end{equation}
for some values $\lambda (s)$ at given $S$ with $k\rightarrow 0$. The values of $S$ are under the requirement $T_c \geq \lambda^{-1/2} n^{2/3}/(A\,\phi_0)$ for $\lambda < 1$. The term 
(\ref{e10}) is singular at $ s\geq S$ because of the singularity in $\xi^{2S}(\lambda)$.  For some values of $\lambda (s)$ at given $S$ the term $G_{CC}$ defines dominantly the asymptotic behaviour $G(k;\lambda)$ at $k\rightarrow 0$, and, consequently, the probability $P(x,\lambda)$ at $x\rightarrow\infty$ (an increasing of the linear size of the BS). The special solution for CC phase may be obtained in the $k$-power form when the infinite series (\ref{e10}) with the integer even power $2S$ of $k$ is replaced by 
 \begin{equation}
\label{e11}
G_{CC}\left [k;\lambda (S)\right ] =  C\left [\lambda (S)\right ] {\vert k\vert}^{r[\lambda(S)]}\cdot Q\left (\vert k\vert \right ),
\end{equation}
where $C(\lambda)$ does not depend on $k$; $Q$ is the amplitude function of $k$. For any $\lambda$ from an open interval $(0,1)$  there will be finite $S$ from the semi-open interval $[0,1)$ when the function $\lambda^{-1} (S)$ can exist as well. Since the fundamental length $\xi(\lambda)$ is finite for $\lambda < 1$ if $s = 1,2,..., S-1$, and it will diverge if $s \geq S$, the series $G_{DM}$ is breaking down to the term of the order $\sim k^{2(S-1)}$. In order to get $r[\lambda(S)]$ in (\ref{e11}), the amplitude $Q\left (\vert k \vert\right )$ has to be clarified first. 
%The $Q\left (\vert k \vert\right )$ 
This amplitude is the associated function of the first order to the power $\gamma$ with the proper (homogeneous) function of the operator $u$ related to similar transformations: $u \,Q\left (\vert k \vert\right ) = Q\left ( a\cdot\vert k \vert\right )$. For any $a > 1$ one has 
 \begin{equation}
\label{e12}
Q\left ( a\cdot\vert k \vert\right ) = a^\gamma Q\left (\vert k \vert\right ) + a^\gamma\, \ln (a)\, Q_0\left (\vert k \vert\right ),
\end{equation}
where $Q_0\left (\vert k \vert\right )$ is the homogeneous function of the power $\gamma$. The result for $r[\lambda(S)]$ is 
 \begin{equation}
\label{e13}
r\left [\lambda(S)\right ] = \frac{1}{\ln(a)}\cdot \ln\frac{Q\left (\vert k \vert\right )}{\lambda (S)\, a^\gamma \left [Q\left (\vert k \vert\right ) + \ln (a)\cdot Q_0\left (\vert k \vert\right )\right ]}.
\end{equation}
The power rank $r$ is trivial at the critical temperature $T_c\sim O\left [n^{2/3}/(A\phi_0)\right]$ with $r_\epsilon \left [\lambda(S)\right ] \simeq \epsilon^{-1} \ln \left [\lambda(S)\right]$, where $\epsilon = \left (A\phi_0/n^{2/3}\right )\cdot \left (1 - T_0/T_c \right)$ for small deviation of the temperature $T_0$ around $T_c$. The coefficient $C\left [\lambda (S)\right ]$ in (\ref{e11}) is 
 \begin{equation}
\label{e14}
C\left [\lambda (S)\right ]\sim \xi^{r[\lambda(S)]} \left [\lambda(S)\right ].
\end{equation}
The functional equation $u \,Q\left (\vert k \vert\right ) = Q\left ( a\cdot\vert k \vert\right )$ admits the special case for the amplitude $Q\left (\vert k \vert\right)$: $Q = const$. On the other hand, $Q\left (\vert k \vert\right)$ may be the smoothly changing function, e.g., the logarithmic-periodic function in $\ln \vert k\vert$ with period $\ln a > 0$. In this case, the solution for an oscillatory function $Q\left (\vert k \vert\right)$ is the infinite series
\begin{equation}
\label{e15}
Q\left (\vert k \vert\right)\sim \frac{1}{\ln a}\, \sum_{m = 0}^\infty b_m\,\cos \left [ 2\pi m\frac{\ln \vert k\vert }{\ln a} + \varphi_m\right ],
\end{equation}
%where the replacements $\vert k\vert \rightarrow  log \left (\vert k\vert \right )$, $a\cdot \vert k\vert \rightarrow log (a) + log \left (\vert k\vert \right )$ have been carried out; 
%which is periodic in $\ln \vert k\vert$ with period $\ln a$; 
where $b_m$ and $\varphi_m$ are the coefficients and the phases, respectively, which are not important for our consideration. 
%The equation for the probability to find the DM in the free state and in the CC phase is 
%\begin{equation}
%\label{e16}
%G(k;\lambda) = G_{DM} (k;\lambda) + \hat G_{CC} (a\cdot k;\lambda),
%\end{equation}
%where $\hat G_{CC} (a\cdot k;\lambda)\sim \xi^{r[\lambda(S)]} \left [\lambda(S)\right ]\cdot {\vert k\vert}^{r[\lambda(S)]}\cdot Q \left (\vert k\vert\right ) $, and $G_{DM} (k;\lambda)$ is in (\ref{e9}).
\section{BS growth}
Now, let us consider the BS growth in terms of the time evolution of the probability distribution $\Pi_n (x;\lambda)$ (\ref{e2}) for the position of the DM "lump" inside the BS after $n$ steps. If the latter occur at equal time $\tau$, then in the limit $\tau\rightarrow 0$ one has the continuous distribution for $\Pi (x,t;\lambda)$ at time $t$ after the random walk begins:
\begin{equation}
\label{e17}
%\partial_t \Pi (x,t;\lambda) = \lim_{\tau\rightarrow 0}\frac{1}{\tau} \int_{-\infty}^{+\infty} \left [P(x-x^\prime;\lambda) - \delta (x-x^\prime)\right ] \Pi (x^\prime, t;\lambda) d x^\prime.
\partial_t \Pi (x,t;\lambda) = \lim_{\tau\rightarrow 0}\frac{1}{\tau} \int_{-\infty}^{+\infty} \left [P(\tilde x) - \delta (\tilde x)\right ] \Pi (x^\prime, t;\lambda) d x^\prime,
\end{equation}
where $\tilde x = x- x^\prime$.
The finite result for (\ref{e17}) is achieved if we put $\Delta\rightarrow 0$ in $P(x;\lambda)$ (\ref{e2}). The Fourier transformation against (\ref{e17}) is useful to ensure the finite result in $k$-space,
 \begin{equation}
\label{e18}
\partial_t g (k,t;\lambda) = \lim_{\tau,\Delta\rightarrow 0}\left \{\frac{1}{\tau} \left [G(\Delta \bar k;\lambda) -1\right ]\right \} g (k,t;\lambda).
\end{equation}
Having in mind that $\lambda < 1$ and $a > 1$, the BS grows with the probability to form the DM "lump" within the power-law cascades (see also [18])
%the time evolution of the probability to form the DM "lump" is given by the power law expression [15]
 \begin{equation}
\label{e19}
\partial_t g (k,t;\lambda) = - B(\varepsilon) {\vert \bar k\vert }^{\varepsilon} g (k,t;\lambda), 
\end{equation}
where $\bar k = k/\Lambda_{cut}$, $\varepsilon = (1- \lambda)/(a -1)$.
 The function $B(\varepsilon)$ depends on the physical meaning  both  $\lambda$ and $a$,
$$ B(\varepsilon) =\frac{\pi}{2}\,\frac{1}{\Gamma \left (\varepsilon \right )\, \sin \left (\pi\varepsilon/ 2\right )}\cdot  \lim_{\tau,\Delta\rightarrow 0} \frac{1}{\tau}\Delta ^{\varepsilon},\,\,\,  0 < \varepsilon < 2 .$$
%The BS growth stops when $a\rightarrow 1$ (the scalar DM field mass $m\sim \Delta_\star \phi_0$), or $\lambda\rightarrow 1$ ($\rho_\star \simeq \rho_\odot$ and/or in the %case of small correlation area, $C_2 (q\rightarrow \infty) \simeq 1$, $\langle N\rangle\rightarrow\infty$). 

%Based on the assumptions, that $\lambda = 1 - \alpha\Delta +o(\Delta)$ and $a = 1 + \beta\Delta +o(\Delta)$, where $0 < \alpha < 2\beta$, the time evolution of the probability %to form the DM "lump" is given by the power law [15]
 %\begin{equation}
%\label{e19}
%\partial_t g (k,t;\lambda) = - B(\alpha, \beta) {\vert \bar k\vert }^{\alpha/\beta} g (k,t;\lambda), 
%\end{equation}
%where $\bar k = k/\Lambda_{cut}$; $B(\alpha, \beta)$ depends on $\alpha$ and $\beta$ with respect to physical meaning of $\lambda$ and $a$,
%$$ B(\alpha, \beta) =\frac{\pi}{2}\,\frac{1}{\Gamma \left (\alpha/\beta \right )\, \sin \left (\pi\alpha/2\beta\right )}\,  \lim_{\tau,\Delta\rightarrow 0} \frac{1}{\tau}\Delta ^{\alpha/\beta}.$$

\section{The special cases for $\lambda$}
The key variable in the model considered is the random fluctuation weight $\lambda$.  We consider two special cases for $\lambda$  to test the "cross-over" from  DM to  CC phase.

Case A. Let $\lambda$ is characterised by the relative energy density of the BS, $\lambda = \rho_\star/\rho_\odot$, where the DM term (\ref{e7}) is finite if $\lambda\rightarrow 0$ and the CC mode (\ref{e8}) is also finite for $\lambda\rightarrow a^{-2}$ at $\phi_0 \neq 0$. Actually, in the interval $ 0 < \rho_\star/\rho_\odot < 1$, the "cross-over" may happen many times as a cascade depending on $T_c$ and $n$ with boundary condition $\rho_\odot < a^2 \rho_\star$. The singular term (\ref{e10}) is accompanied by the lower bound on $T_c \geq \sqrt{\rho_\odot/\rho_\star}\, n^{2/3}/(A\phi_0)$ with $\rho_\star \neq 0$. From the differential Eq. (\ref{e19}) we find the range for $\Delta_\star r_\star$, when the BS is "clustered" by the DM "lumps",
$$ \frac{A\, T_c}{n^{2/3}} >  \Delta_\star r_\star > \frac{3}{2\,\phi_0} \left (1 - \frac{\rho_\star}{3\rho_\odot}\right ).$$
The lower bound for the scale $\Delta_\star r_\star$ (the BS star radius if $\Delta_\star\sim\sqrt{g}$) is given by the scalar DM condensate $\sim \phi_0^{-1}$ only, when the BS energy density  $\rho_\star$ is negligible compared to that of the $\rho_\odot$. 

%For every $z=k/(\pi \Lambda_{cut})$ where random walks begin inside the star from the local point $z$ to smearing point $z+\Delta z$ in the sense of Weierstass function (see Sec.2), one has the lower %bound on  the BS radius $R_{low}^\star > \phi_0^{-1} \rho_\odot/\rho_\star$. For $\rho_\star/\rho_\odot << 10^{-10}$ and $\lambda_4 = 6\cdot 10^9$ [8] one has the BS radius  $R_\star >>>R_{low}%^\star\sim 10^{-7}$ cm  decreasing from $10^7$ km to 10 km based on the CMS data [18] on the DM interactions through the Higgs portal for the $\phi$ boson mass range $[0.1 GeV, m_h/2]$.
Case B. The sources (with the characteristic size $\sim L_{st}$) of the DM particles correlated according to CBE may be useful to present the $\lambda$ in the form $\lambda = C_2 (q)\eta (N) -1$ 
%($\bar C_2 (q) = C_2 (q)\eta (N)$)
, where the DM term (\ref{e7}) is finite if $\lambda\rightarrow 0$ at $q\rightarrow\infty$ and $N\rightarrow \infty$. 
In the CC mode,
% the (\ref{e8}) is also finite where the lower bound  
%$$ \bar C_2 (q) > 1 + \left (\frac{n^{2/3}}{A\,\phi_0\, T_c}\right )^2 $$
%means the correlations at large $q$ (small $L_{st}$) with the coherence function $\lambda_{ch}\simeq 1$ (the fully chaotic phase where the "cross-over" can occur), and $n\rightarrow 0$ at %$T_c\rightarrow\infty$. 
the following space-momentum relation can be found relevant to (\ref{e10}) 
$$ L_{st}^2\cdot q^2 \leq \ln \lambda_{ch} (\nu) - \ln \left [\eta^{-2}(N) \left (1 + a^{-2}\right ) -1\right ],$$
where 
$$ T_c > \frac{n^{2/3}}{A\,\phi_0} \frac{1}{\sqrt{ 2\, \eta^2 (N) -1}} $$
with $\lambda_{ch} \leq 1$. Actually, $T_c > n^{2/3}/(A\,\phi_0)$ for large enough multiplicities $N$. The lower bound for $\Delta_\star r_\star $ is given by
$ \Delta_\star r_\star > \phi_0^{-1} \left\{ 2 -\left [1 - \epsilon_2 (q)\right ] \eta (N)\right \}, $
where $\epsilon_2 (q) = 1 - (1/2) C_2(q).$  It is easy to find that the special cases A and B for $\lambda$  will be simultaneously acceptable if
 \begin{equation}
\label{e20}
\frac{\rho_\star}{\rho_\odot} = 2\left [ 1 - \epsilon_2 (q)\right ] \eta (N) -1.
\end{equation}
Actually, the (\ref{e20}) is the energy density instability in the spatially uniform state with CBE, where the result tends to zero at $q\rightarrow\infty$ 
if $\langle N\rangle\rightarrow\infty $. Thus, even the Universe were dominated by the scalar DM fields, their presence would weakly affect the CBE. Taking into account indefiniteness in the Universe related to  (\ref{e20}), one may expect essentially larger energy density of the star, $\rho_\star \simeq \rho_\odot$ in the large space area of particle correlations for large enough average multiplicities. 
The BS growth stops when $a\rightarrow 1$ (the scalar DM field mass $m\sim \Delta_\star \phi_0$), or $\lambda\rightarrow 1$ ($\rho_\star \simeq \rho_\odot$ and/or in the case of large enough correlation area, $C_2 (q\rightarrow 0) \leq 2$). 
In this case, observational astrophysical restrictions for the spectrum of the direct photons (or dark photons) due to decays of the BS (through the scalar DM decays) would enable the restriction of the scalar DM dominance stages in the early Universe.

In the conclusion of this section, the attention is concentrated to the BS radius. For every $z=k/(\pi \Lambda_{cut})$ where random walks begin inside the star from the local point $z$ to smearing point $z+\Delta z$ in the sense of Weierstass function (see Sec.2), one has the lower bound on  the quantity $R_{low}\sim \Delta_\star r_\star  > \phi_0^{-1} \rho_\odot/\rho_\star$. For $\rho_\star/\rho_\odot < 10^{-10}$ and $\lambda_4 = 6\cdot 10^9$ [9] one has the BS radius  $R_\star\sim \sqrt g\, m_\phi^{-1} >>> R_{low}\sim 10^{-7}$ cm, where $R_\star$  being down  from $10^5$ km to 10 km based on the CMS data [19] on the DM interactions through the Higgs portal for the $\phi$ boson mass range $[0.1 GeV, m_h/2]$. In case when $\lambda$ is governed by the CBE dynamics with $q$, $\lambda_{ch}$ and $\langle N\rangle$, the lower bound $R_{low} > \phi_0^{-1}\,{\left [C_2 (q, \lambda_{ch}) \eta(N) - 1\right ]}^{-1}$ from which one easily finds that BS size may be infinitely large in case of huge number of particles, $\langle N\rangle\rightarrow\infty$, when the correlation domain is rather small ($q\rightarrow\infty$). On the other hand, at small values of $q$ (large enough stochastic scale $L_{st}$), $R_{low}\sim \phi_0^{-1}\sim 10^{-16}$ cm which may correspond to the characteristic minimal size of the DM cluster, the scalar DM "lump".

\section{Conclusion} 
In conclusion, we studied the dynamics in the formation of the BS through the "cross-over" between the free scalar DM and the CC phase in one-dimensional model of the RFW. The main inputs are the random fluctuating weight $\lambda < 1$, the fundamental fluctuating length $\xi (\lambda)$ and the parameter $a >1$ to the spatial separation between the DM "lumps" inside the BS. The solution to the functional equation for the characteristic probability to find the scalar DM in the free state and in the CC phase is obtained. 
The time evolution of the probability to form the DM "lump" (growth of the BS) is presented with the power-law cascades (\ref{e19}).
The special cases for $\lambda$ in terms of the energy densities of the BS, $\rho_\star$, and the DM, $\rho_\odot$,  as well as the CBE $C_2$ - function are considered. The lower bounds on the critical temperature at the "cross-over" depending on  $\sim \rho_\odot/\rho_\star$, the average multiplicities $N$ of the scalar DM particles and the parameters $\lambda_{ch}$, $\nu$, related to the chaotically behaviour of the particles, are obtained.   We have shown how self-similar DM "lumps" can arise in $ D=1$ spatial dimension within the RFW in BS under stochastic and chaotic processes defined by $\lambda_{ch} (\nu)$. Other random walks that have non-self-similar clusters of the "lumps" in $D >1$ spatial dimensions, can be studied in a similar manner. 

%\section{References} 


\begin{thebibliography}{9}
\bibitem{1}
R. Ruffini and S. Bonazzola, Phys. Rev. 187, 1767 (1969).
\bibitem{2}
I.I. Tkachev, Sov. Astron. Lett. 12, 305 (1986).
\bibitem{3}
M. Colpi, S.L. Shapiro, and I. Wasserman,  Phys. Rev. Lett. 57, 2485 (1986).
\bibitem{4}
A. Ahmed, A. Mariotti and S. Najjari,  JHEP 05, 093 (2020).
\bibitem{5}
K. Griest and M. Kamionkowski,  Phys. Rev. Lett. 64, 615 (1990).
\bibitem{6}
I. Baldes and K. Petraki,  J. Cosmol. Astropart. Phys. 09, 028 (2017).
\bibitem{7}
J. Smirnov and J.F. Beacom,  Phys. Rev. D100, 043029 (2019).
\bibitem{8}
G.A. Kozlov,  J. Math. Phys. 42, 4749 (2001).
\bibitem{9}
G.A. Kozlov,  Mod. Phys. Lett. A 38, 2350005 (2023).
\bibitem{10}
D.G. Levkov, A.G. Panin, and I.I. Tkachev, Phys. Rev. Lett. 121, 151301 (2018).
\bibitem{11}
G.A. Kozlov,  Phys. Part. Nucl. Lett.  6, 97 (2009).
%\bibitem{11}
%G.A. Kozlov, {\it ....} {....}, .... (....).
%\bibitem{11}
%D. Kharzeev, {\it ....} {....}, .... (....).
%\bibitem{12}
%A.A. Andrianov, {\it ....} {....}, .... (....).
\bibitem{12}
A.S. Dmitriev et al., Phys. Rev. Lett. 132, 091001 (2024). 
\bibitem{13}
J. Eby et al., J. High Energy Phys. 02, 028 (2016). 
\bibitem{14}
J. Garcia-Bellido, J. Rubio, M. Shaposhnikov, and D. Zenha,  Phys. Rev. D84, 123504 (2011).  
\bibitem{15}
M.Yu. Khlopov, B.A. Malomed and Ya.B. Zeldovich,   Mon. Not. R. astr. Soc. 215, 575 (1985).  
\bibitem{16}
M.P. Hertzberg et al.,  Phys. Rev. D 103, 023536 (2021).
\bibitem{17}
G.H. Hardy,  Trans. Am. Math. Soc. 17, 301 (1916).
\bibitem{18}
B.D. Hughes, M.F. Shlesinger, and E.W. Montroll,  Proc. Natl. Acad. Sci. USA 78, 3287 (1981).
\bibitem{19}
CMS Collaboration,  Eur. Phys. J. C83, 933 (2023). 



\end{thebibliography}
\end{document}